\documentclass[pra,twocolumn,preprintnumbers,amsmath,amssymb]{revtex4}

\usepackage{graphicx}
\usepackage{dcolumn}
\usepackage{bm}
\usepackage{epsfig}
\usepackage[dvipsnames]{color}
\usepackage{epstopdf} 

\newcommand{\beq}{\begin{eqnarray}}
\newcommand{\eeq}{\end{eqnarray}}

\renewcommand{\Re}{\operatorname{Re}}

\newcommand{\figurewidth}{\columnwidth}
\input xy 
\xyoption{all}
\definecolor{myback}{rgb}{1,.964,.8}

\begin{document}

\title{Solving the Graph Isomorphism Problem with a Quantum Annealer \\
}
\author{Itay Hen}
\email{itayhe@physics.ucsc.edu}
\affiliation{Department of Physics, University of California, Santa Cruz, California 95064, USA}
\author{A.~P.~Young}
\affiliation{Department of Physics, University of California, Santa Cruz, California 95064, USA}


\begin{abstract}
We propose a novel method using a quantum annealer -- an analog quantum computer based
on the principles of quantum adiabatic evolution -- to solve the Graph
Isomorphism problem, in which one has to determine whether two graphs are
isomorphic (i.e., can be transformed into each other simply by a relabeling of
the vertices). We demonstrate the capabilities of the method by analyzing
several types of graph families, focusing on graphs with particularly high
symmetry called strongly regular graphs (SRG's). 
We also show that our method is applicable, within certain limitations, to
currently available quantum hardware such as ``D-Wave One''.
\end{abstract}

\pacs{03.67.Ac,02.10.Ox,03.67.Lx}
\keywords{Graph Isomorphism Problem, Quantum Adiabatic Algorithm, Quantum Adiabatic Computation} 
\maketitle
\section{\label{intro}Introduction}

Theoretical research on quantum computing is motivated by the exciting
possibility that quantum computers may be able to perform certain tasks faster
than classical computers. In recent years first steps have been taken towards
the goal of experimentally realizing these computational advantages.  However,
to date the largest experimental implementations of scientifically meaningful
quantum algorithms have used just a handful of qubits~ \cite{vandersypen:01,
gaitan:12}, the reason being the tremendous technological challenges (the most
crucial of which is overcoming quantum decoherence) that need to be defeated
before a successful implementation of any solid-state quantum computer.


Of the several quantum computing paradigms that have been proposed, a
potentially promising substitute for the `standard' circuit-based quantum
computer is the quantum annealer~\cite{kadawoki:98,farhi:01b}, which is now on
the cusp~\cite{gaitan:12} of being able to run small-scale computing procedures on
actual quantum annealing hardware.  Quantum annealing machines are analog
quantum computational devices designed to solve discrete combinatorial
optimization problems using properties of quantum adiabatic evolution.  They
are based on a general approach widely known as the Quantum Adiabatic
Algorithm (QAA), which was proposed by Farhi {\it et al.}~\cite{farhi_long:01}
about a decade ago as a method for solving a \textit{broad range} of
optimization problems using a quantum computer. 

Recently, a quantum annealing machine, based on liquid crystal nuclear
magnetic resonance has been reported to successfully factor the number $143$
using four spin-qubits~\cite{xu:11}.  A few months later, ``D-Wave One'', a
$128$-qubit machine based on super-conducting qubits~\cite{johnson:11} experimentally demonstrated the ability to compute two-color Ramsey numbers \cite{gaitan:11,gaitan:12} making it the largest experimental implementation of a
scientifically meaningful quantum algorithm. 
This success provides
an incentive for finding problems 
that could be solved efficiently on a quantum annealer.

In this paper, we hypothesize that a quantum annealer could
solve the Graph Isomorphism (GI) problem, described in detail below. In
particular, we 
hypothesize that the annealer
could distinguish all non-isomorphic graphs by sufficiently precise measurements.
We provide some evidence for this by showing that our
method works for certain graphs which we have been able to study numerically. 
An advantage of our proposal is that it can,
within certain limitations, be implemented using {\it
currently available} quantum annealing machines such as D-Wave One.

The paper is organized as follows. In Sec.~\ref{sec:GI} we describe the Graph
Isomorphism (GI) problem in some detail.  In Sec.~\ref{sec:QAE} we describe
and then discuss the proposed method for solving the GI problem using quantum
adiabatic evolution.  We next present the results of the study in
Sec.~\ref{sec:results} and derive some conclusions in
Sec.~\ref{sec:conclusions}. 

\section{\label{sec:GI}The Graph Isomorphism Problem}

A graph $G = (V,E)$ is a set of vertices $V$, and edges $E$ which are unordered pairs 
of vertices.  A graph is conveniently expressed algebraically as an \emph{adjacency matrix}.
The adjacency matrix $A$ of a graph with $N$ vertices is an $N \times N$ matrix
in the basis of vertex labels, with $A_{ij}=1$ if vertices $i$ and $j$ are
connected by an edge, and zero otherwise. 

The GI problem is stated as follows:
Given two graphs, one must determine whether or not they are isomorphic to
each other, i.e., whether one can be transformed into the other by a
relabeling of the vertices.

In the realm of classical computing, many special cases of GI have been shown
to be solvable in a time that scales as a polynomial of the number of
vertices. However the best {\it general}  algorithm to date runs in time
$\mathcal O \left( a^{ N^{1/2}\log N} \right)$ where $a$ is a
constant~\cite{spielman:96}. It is therefore interesting 
to ask whether a quantum computer could solve this problem efficiently.
An attractive feature of the GI problem is
its resemblance to the problem of integer factoring, the first and
most famous example to date of a quantum algorithm can solve a problem
exponentially faster than the best known classical algorithm \cite{shor:94}.
Like factoring, the common belief is that the GI problem is unlikely to be
NP-complete. 

The GI problem has been attacked by numerous methods inspired by classical as
well as quantum physical systems.  Rudolph~\cite{rudolph:02} mapped the GI
problem onto a system of hard-core atoms, where one atom was used per vertex,
and atoms $i$ and $j$ interacted if vertices $i$ and $j$ were connected by
edges. It was demonstrated that for some pairs of non-isomorphic graphs,
sharing cospectral adjacency matrices does not lead to cospectrality of the
transition matrix between three-particle states produced by the embedded
Hamiltonian.  Gudkov and Nussinov~\cite{gudkov:02} proposed a classical
algorithm to distinguish non-isomorphic graphs by mapping them onto various
physical problems.  Shiau \emph{et al.} proved that the simplest classical
algorithm fails to distinguish some pairs of non-isomorphic graphs and also
proved that continuous-time one-particle quantum random walks cannot
distinguish some non-isomorphic graphs \cite{shiau:05,gamble:10,rudinger:12}.
More recently, it has been found that classical random walks and quantum
random walks can exhibit qualitatively different properties
\cite{aharonov:93,bach:04,solenov:06}. These disparities mean that \textit{in
some cases}
algorithms implemented
by quantum random walkers can be proven to run faster than the fastest
possible classical algorithm
\cite{childs:02,shenvi:03,ambainis:03,ambainis:04,magniez:07,potocek:09,
reitzner:09,douglas:08,emms:06,emms:09}.

\section{\label{sec:QAE}Quantum Adiabatic Evolution and the GI Problem}

To solve the GI problem using quantum adiabatic evolution, we
assign to each graph $G$ the Hamiltonian
$\hat{H}(G)$, where
\beq \label{eq:H}
\hat{H}(G)=(1-s) \hat{H}_d +s \hat{H}_p(G) \,,
\eeq
which depends on a parameter $s$. For $s=0$, $\hat{H}(G)$ is the
standard 'driver' Hamiltonian for QAA algorithms, namely
\begin{equation}
\hat{H}_d=\frac1{2} \sum_i^N \sigma_i^x,
\end{equation}
i.e., a transverse-field Hamiltonian, while for $s=1$, $\hat{H}(G)$ is the 'problem' Hamiltonian
$\hat{H}_p(G)$ which is constructed according to the topology of the graph.
A simple plausible choice for $\hat{H}_p(G)$ is
\begin{equation}
\hat{H}_p(G) =\sum_{\langle i j \rangle  \in G} \sigma_i^z \sigma_j^z \,,
\label{HP}
\end{equation}
i.e., an Ising antiferromagnet on the edges of the graph.
While the driver Hamiltonian is diagonal in the $\prod \sigma^x$ basis, the
problem Hamiltonian is diagonal in the $\prod \sigma^z$ basis. 

The system is first prepared in the ground state of the driver
Hamiltonian $\hat{H}_d$, which is straightforward.
The adiabatic parameter $s$ is then varied
slowly and smoothly with time
from $0$ to $1$, so that the Hamiltonian is
continuously modified from $\hat{H}_d$ to $\hat{H}_p(G)$.  If this process is
done slowly enough, the adiabatic theorem of Quantum Mechanics (see, e.g.,
Refs.~\cite{kato:51} and~\cite{messiah:62}) ensures that the system will stay
close to the ground state of the instantaneous Hamiltonian throughout the
evolution.

The premise of the method we suggest here is that the state of the system
along the adiabatic evolution, i.e., the instantaneous ground state of the
Hamiltonian, Eq.~(\ref{eq:H}), stores enough information to reflect the
complex structure of the graph-dependent problem Hamiltonian $\hat{H}_p(G)$,
and that carefully chosen measurements along the adiabatic path will provide
enough information to
differentiate non-isomorphic graphs. 


The choice of the problem Hamiltonian in Eq.~\eqref{HP} is,
of course, only one possibility.  However it has
the advantages of (i) using Ising spins which are simple to study and to
implement experimentally~\cite{johnson:11,gaitan:12},
and (ii) having antiferromagnetic interactions which, on
highly-connected graphs, tend to have highly frustrated ground-states because
closed paths (along the edges) of odd length make the system a spin
glass~\cite{mezard:01,mezard:03b,zdeborova:09,farhi:12}. Note also, that the
Hamiltonian Eq.~(\ref{eq:H}) is symmetric with respect to flipping all spins. 

It is interesting to compare our approach with those which use
Hamiltonians of quantum random walkers embedded in the graph structure.
In the latter case the walker normally accesses
only low-dimensional subspaces of the Hamiltonian eigenstates, based on
conserved quantities such as number of particles (see, e.g.,
Refs.~\cite{rudolph:02,shiau:05,emms:06,emms:09}), whereas in our approach
there is no such conserved quantity so the instantaneous ground state of the
Hamiltonian presumably reflects the full complexity of the graph.

For each graph, one 
performs multiple runs along the adiabatic path, performing various
measurements at different values of $s$ until sufficient statistics
is gathered.  Since each measurement collapses the state of the system,
non-commuting measurements or measurements corresponding to different $s$
values, require separate runs of the procedure.  Since errors of the
various measured quantities are inversely proportional to square root of the
number of measurements, this number will be determined from the needed
resolution.

In order to make sure that isomorphic graphs are recognized as such, the
measurements must be invariant under a relabeling of the vertices
of the graph (or equivalently the spins in the system).  The most
straightforward such quantity is the Hamiltonian, so the average energy
$E=\langle \hat{H}(G) \rangle$ is a `good' quantity to measure.  Here, $\langle
\cdot \rangle$ indicates the expectation value with respect the state of the
system, i.e., the instantaneous ground-state, at a particular value of $s$.
For the same reason, the classical `diagonal' average energy of the graph
$E_G=\langle \hat{H}_p(G) \rangle$ and $M_x=2 \langle \hat{H}_d \rangle$, the
$x$-magnetization, are also suitable observables.

Many other quantities that respect the topology of the graph
and are invariant under any relabeling of the vertices could be
measured.  Some may prove to have better distinguishing capabilities than others,
depending on the manner in which they `tap' into the complexity of the
structure of the graph in question.  

We find that a `good' quantity to measure is the
spin-glass order parameter, which we shall denote as $Q_2$ and define by
\beq \label{eq:q2}
Q_2= \left({1\over N(N-1)}\,  \sum_{i\ne j} \langle \sigma_i^z \sigma_j^z
\rangle^2 \right)^{1/2}  \,.
\eeq
Generalizing the above expression, the quantities
\beq
Q_{2n} = \frac1{N^{2n}} \sum_{i_1,i_2,\ldots,i_{2n}} \langle
\sigma_{i_1}^z \sigma_{i_2}^z \ldots  \sigma_{i_{2n}}^z\rangle^2 \,,
\eeq
where \hbox{$n=2,3,4 \ldots$}, also serve as distinguishing operators. 
Analogously, other types of operators that might work are of the form:
\beq
Q_2' = \frac1{N^2} \sum_{i,j} \langle \hat{H}_p(G) \sigma_i^z \sigma_j^z \rangle^2 \,.
\eeq
Each such measurable quantity, if measured sufficiently accurately, for
different values of the adiabatic parameter $s$, serves as an additional
`dimension' of differentiability of the non-isomorphic graphs.  Moreover,
gathering statistics of these quantities for different values of $s$ accesses
indirectly the entire spectrum of $\hat{H}_p(G)$ (unlike the situation in
classical or quantum random walks where the walkers are restricted to only a small subspace of the Hamiltonian eigenstates).

In Sec.~\ref{sec:results} we will consider several sets of non-isomorphic
graphs, showing that, in all cases,
non-isomorphic graphs are
distinguished by accurate measurements of carefully chosen observables
along the adiabatic path. In fact, we shall see that in most cases studying
the limit $s \to 1$ suffices.

\section{\label{sec:results}Results}

We now present numerical and semi-analytical results for
several types of graph families that are
known to be hard to distinguish. Since the annealing process requires that the
temperature of the system be well below the excitation gap of the system, 
we shall work at zero temperature.

For the smaller graphs that we study ($N \leq 25$ vertices), we use
exact-diagonalization or conjugate-gradient based minimization techniques.
Both techniques are restricted to quite small values of $N$ because the size of the
Hilbert space grows as $2^N$.  The specific variation of conjugate-gradient method
that we developed for this study is discussed in Appendix~\ref{app:cgm}.

For graphs of more than $25$ vertices, quantum Monte-Carlo techniques would
usually be the method of choice for the accurate measurements of the various
quantities. However, these were found by us to be rather inefficient in the
interesting regions where the value of the adiabatic parameter is close to
$1$.  We have been able to study sizes \textit{a little} larger than $25$ (up
to $29$) by leading-order degenerate perturbation theory about the limit $s =
1$.  This calculation involves finding the subspace of ground states of
$\hat{H}_p(G)$, which requires one to first calculate the energies of the
$2^N$ `classical' states followed by further manipulation on the subspace
spanned by the ground-state eigenvectors. The main shortcoming of the method
is that it only produces results in the $s \to 1$ limit.  Fortunately, we find
that in most cases this is sufficient.

The measured quantities we focus on are the total (average) energy, the
spin-glass order parameter $Q_2$ given in Eq.~(\ref{eq:q2}) and the
$x$-magnetization $M_x=2 \langle \hat{H}_d \rangle$. We find that in
most cases these three quantities are sufficient to distinguish all tested
non-isomorphic graphs, though these measurements may be augmented by
measuring other observables, as discussed in the previous section.

\subsection{\label{sec:srgs}Strongly regular graphs}

\label{sec:SRG}
The main results of this paper are for strongly regular
graphs (SRGs), a class of graphs, subsets of which are known to be difficult
to distinguish~\cite{gamble:10}.  An SRG is a graph in which (i) all vertices
have the same degree, (ii) each pair of neighboring vertices has the same
number of shared neighbors, and (iii) each pair of non-neighboring vertices has
the same number of shared neighbors. This definition permits SRGs to be
categorized into families by four integers $(N,k,\lambda, \mu)$, each of which
might contain many non-isomorphic members. Here, $N$ is the number of vertices
in each graph, $k$ is the degree of each vertex (k-regularity), $\lambda$ is
the number of common neighbors shared by each pair of adjacent vertices, and
$\mu$ is the number of common neighbors shared by each pair of non-adjacent
vertices.

Using the stringent constraints placed on SRGs, one can show that, for any
SRG, the spectrum of the adjacency matrix
has only three distinct values~\cite{godsil:01}:
$\lambda_0 = -k$, which is non-degenerate, and
$\lambda_{1,2} = -\frac{1}{2} \left( \lambda - \mu \pm \sqrt{N} \right)$,
which are both highly degenerate. 
Both the value and degeneracy of these eigenvalues depend only on the family
parameters, so within a particular SRG family, all graphs are cospectral
\cite{godsil:01}.  These highly degenerate spectra are one reason why
distinguishing non-isomorphic SRGs is difficult.

While there exist SRG families with only one non-isomorphic member
\cite{weisstein}, we concern ourselves with families that have more than one
non-isomorphic graph. Using combinatorial techniques \cite{spence,mckay},
tables of complete
and partial families of SRGs have been
tabulated and we will use those tables to select graphs for our study.  It
should be noted that for every family of non-isomorphic SRGs there exists a
complementary family of the same size $N$ and the same number of members,
which is obtained by interchanging edges and non-edges. 

\begin{figure}
\begin{center}
\includegraphics[width=\figurewidth]{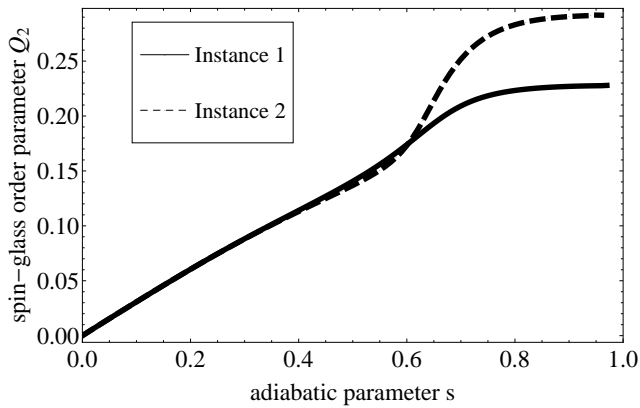}
\caption{Spin-glass order
parameter $Q_2$ in the ground state
for the two non-isomorphic strongly-regular graphs (SRGs) on
$N=16$ vertices. As the figure indicates, the two graphs show different $Q_2$
values starting from $s\simeq 0.4$.\label{fig:n16qs}}
\vspace{-.7cm}
\end{center}
\end{figure}

The smallest family of non-isomorphic SRGs that have more than one member is
that of the $N=16$ vertices with signature $(16,6,2,2)$, 
which contains two members (we shall not address here the complementary
families of graphs which are distinguished in much the same way). The two
graphs are immediately distinguished by looking for example at the spin-glass
order parameter $Q_2$ as a function of the adiabatic parameter $s$. The $Q_2$
values of the two graphs (at zero temperature) are plotted as a function of
$s$ in Fig.~\ref{fig:n16qs}.  In this $N=16$ case, both graphs have the same
`classical' ground-state energy of $E_G=\langle \hat{H}_p(G) \rangle=-16$
albeit with different degeneracies, namely $21$ and $45$~\cite{bit-flip}.

\begin{figure}
\begin{center}
\includegraphics[width=\figurewidth]{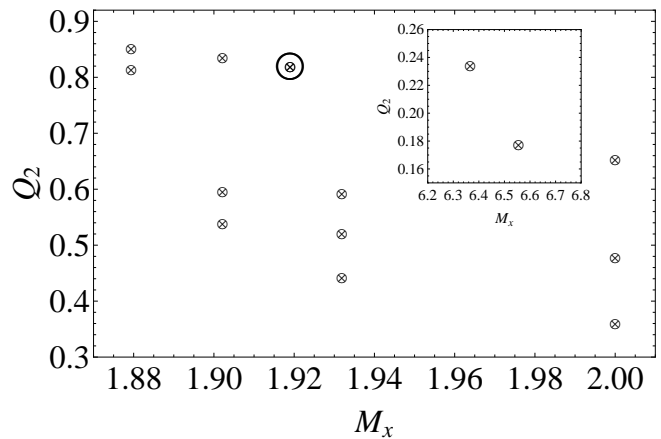}
\caption{Scatterplot in the $M_x-Q_2$ plane of the $15$ strongly regular graphs
with $N=25$ vertices in the limit $s\to1$. 
The horizontal axis is the magnetization along the $x$-direction and the vertical axis is
the spin-glass order parameter $Q_2$.
The circled data
point corresponds to the only two graphs which are not distinguished by values
of  $Q_2$ and $M_x$ in the $s\to1$ limit.
These two
instances are, nonetheless, distinguished by measurements at $s < 1$ (see text). 
The inset shows the $M_x-Q_2$ values of two of the graphs that lie outside of the
region shown in the main panel.
\label{fig:n25}}
\vspace{-0.7cm}
\end{center}
\end{figure}

The next family of SRGs that contains more than one member is that with $N=25$
vertices.  This family has $15$ distinct graphs with signature $(25,12,5,6)$.
We find that looking
at the values of $Q_2$ and $M_x$ in the ground state
in the limit of $s\to1$ using first-order degenerate perturbation
theory suffices to distinguish between all but two of the graphs (the latter
two share the same $Q_2$ and $M_x$ values in this limit).  The results are shown
in Fig.~\ref{fig:n25}, which is a scatterplot of $Q_2$ vs $M_x$ for $s \to 1$ for all 15
non-isomorphic graphs.  In this limit, most instances have the same
ground-state energy of $E_G=-34$ except for two instances which have a slightly
higher energy of $E_G=-30$.  These correspond to the two points in the inset
of Fig.~\ref{fig:n25}.  In addition, the circled data point in the figure
corresponds to the $(M_x,Q_2)$ value of the two graphs which are not
distinguished in the
$s\to1$ limit.  However, for values of $s$ \textit{less} than 1, 
exact diagonalization 
reveals a clear distinction
between the two graphs. For example, for $s=0.73$ the 
values of $Q_2$ in the ground state are $0.57914$ and $0.443423$.

\begin{figure}
\begin{center}
\includegraphics[width=\figurewidth]{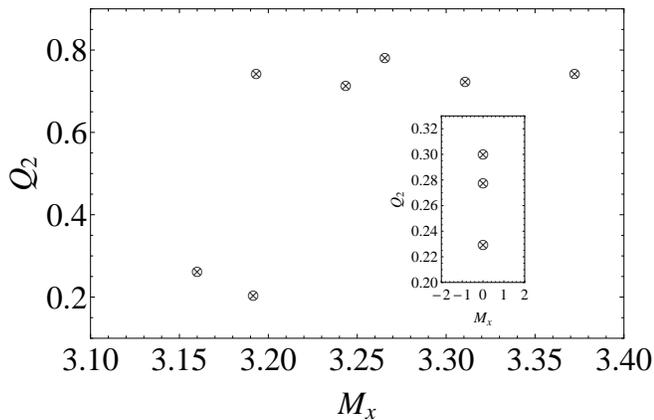}
\caption{Scatterplot in the $M_x-Q_2$ plane of the $10$ strongly regular graphs
with $N=26$ vertices and signature $(26,10,3,4)$, in the $s \to 1$ limit.  As the figure indicates,
the annealer distinguishes all graphs in the family.\label{fig:n26}}
\vspace{-0.7cm}
\end{center}
\end{figure}

Perturbation theory-based analysis in the $s\to 1$ limit also allows us to study
families of graphs with $N=26, 28$ and $29$ vertices.  The degenerate
perturbation-theory analysis of the set of $10$ SRGs with $N=26$ and signature
$(26,10,3,4)$ are shown in Fig.~\ref{fig:n26} which is a scatterplot in the
$M_x-Q_2$ plane. Here, 7 of the instances were found to have $E_G=-34$ (with different degeneracies, all of them around $\sim 1000$) and required first-order perturbation theory, whereas the remaining 3 instances 
had $E_G=-38$ (with degeneracies $20,20$ and $60$), and the leading-order perturbation analysis is of degree 6. 
Fig.~\ref{fig:n26} shows that all 10 members of the family 
are distinguished in that plane. The inset shows data points that lie
outside of the range presented in the main panel.

For $N=28$, the family $(28,12,6,4)$ contains $4$ distinct graphs. There too,
we find that it is sufficient to look at the $Q_2$ values and average total energy
in the $s\to1$ limit.
For three of the four graphs that have the ground-state energy of $E_G=-28$ (and with different degeneracies, all of them around $\sim 6000$)
the $Q_2$ values are $0.137461$, $0.141957$ and $0.132883$.  The fourth graph
has a different energy of $E_G=-24$ (with a degeneracy of $972265$). The $Q_2$
value has not been calculated for this graph. 

\begin{figure}
\begin{center}
\includegraphics[width=\figurewidth]{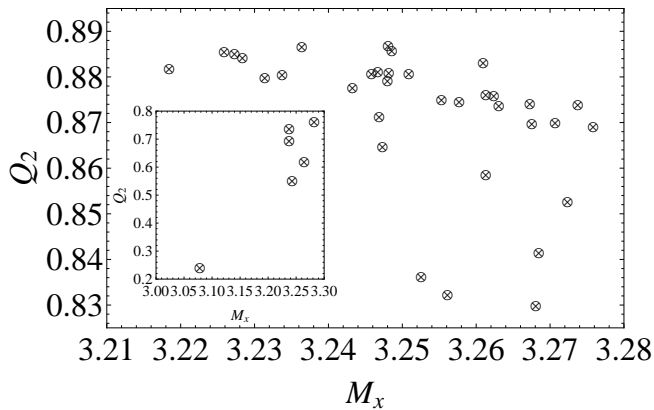}
\caption{Scatterplot in the $M_x-Q_2$ plane of the $41$ strongly regular graphs with
$N=29$ vertices, in the $s \to 1$ limit. 
As the figure indicates, the annealer distinguishes all graphs in the family. The inset shows data points
that are out of the range presented in the main panel.\label{fig:n29} }
\vspace{-0.7cm}
\end{center}
\end{figure}

The largest size of graphs that we deal with analytically using
perturbation theory is the family of $41$ graphs having $N=29$ vertices and
signature $(29,14,6,7)$.  A scatterplot in the $M_x-Q_2$ plane for this family
is given in Fig.~\ref{fig:n29}.  While some of the values shown in the
scatterplot lie close to one another, all pairs $(M_x,Q_2)$ are distinct
(and can be further differentiated by measurements of other observables) so
all non-isomorphic graphs of this family can be distinguished.
Here, all members were found to have the same energy of $E_G=-41$
and with different degeneracies, all of them around $\sim 3000$. 

\subsection{\label{sec:nonsrg}Other pairs of graphs}
Strongly-regular graphs are only one of many classes of graphs that
are considered difficult to distinguish.  Here, we discuss two regular (rather
than strongly-regular) pairs of graphs that are known to be difficult to
distinguish (and were, in fact, constructed to be so). 
The first pair, which we denote here as ($G_1$,$G_2$) was first
considered by Emms {\it et al.} \cite{emms:06,emms:09} (the reader is referred
Appendix \ref{app:nonsrg} for the adjacency matrices of the two graphs).  The
two graphs are regular having 14 vertices.  They were given as an example of
non isomorphic graphs that can not be distinguished by certain quantum random walkers, although it should be noted that other more recent quantum random walkers seem to be able to distinguish between the graphs~\cite{kennyPrivate}.  

\begin{figure}
\begin{center}
\includegraphics[width=\figurewidth]{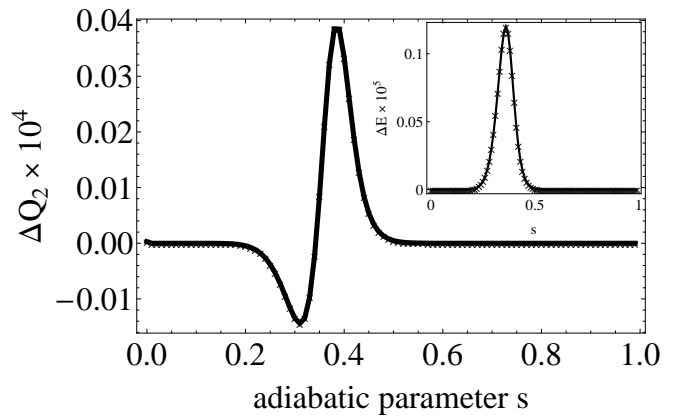}
\caption{ The difference in $Q_2$ values, $\Delta Q_2$ (main panel) and energies (inset), for the two graphs $G_1$ and $G_2$ with $N=14$ that certain quantum walkers cannot distinguish \cite{emms:06,emms:09}.\label{fig:deltaQ14}}
\vspace{-0.7cm}
\end{center}
\end{figure}

While having almost identical adjacency matrices
and therefore also almost identical energies and $Q_2$ values for $s \to 1$, the
two graphs can still be
distinguished by our quantum annealer, at least
within a small region of $s$.  This can be seen by looking at
differences in $Q_2$ values (and smaller but visible differences in the energy).
These are shown in Fig.~\ref{fig:deltaQ14}. Interestingly, this region precisely corresponds to the
quantum phase transition normally seen in quantum adiabatic computing
procedures that is characterized by a small gap~\cite{young:08,young:10}.
This region is therefore the usual `bottleneck' of the Quantum Adiabatic
Algorithm.

\begin{figure}
\begin{center}
\includegraphics[width=\figurewidth]{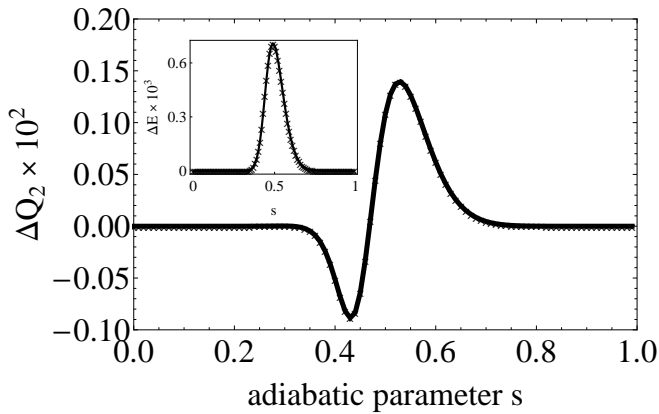}
\caption{The difference in $Q_2$ values, $\Delta Q_2$ (main panel) and energies (inset), for the two graphs $G_3$ and
$G_4$ with $N=16$ that certain quantum walkers cannot distinguish \cite{guo:10}.\label{fig:deltaQ16}}
\vspace{-0.7cm}
\end{center}
\end{figure}

We also considered another pair of non-isomorphic regular graphs on 16 vertices, denoted here
by $G_3$ and $G_4$ (the reader is referred Appendix \ref{app:nonsrg} for the
adjacency matrices of the two graphs), that are known to be difficult to
distinguish~\cite{godsil:01,guo:10}.  
Similarly to the previous pair ($G_1$,$G_2$), the pair ($G_3$,$G_4$) can 
be distinguished by looking at the $Q_2$ and energy differences as a function of $s$, see Fig.~\ref{fig:deltaQ16}. 

Interestingly, if one inspects the spectra of the `diagonal' Hamiltonians $\hat{H}_p(G)$ of the two pairs of graphs, one finds that each pair shares the exact same spectrum, meaning that while `classically' it is impossible to distinguish between the graphs within each pair, extending the analysis to the `quantum region', by adding a non-commuting term (the driver Hamiltonian) enables the crucial missing differentiation.

\section{\label{sec:conclusions}Conclusions and Future Research}


The results we presented here support a conjecture that the Quantum Adiabatic
prescription can differentiate between all non-isomorphic graphs, given an
appropriate choice of problem and driver Hamiltonians.  This conjecture needs
to be tested more thoroughly, both theoretically and also by 
experiments on real quantum annealers.

In future studies, it would be desirable to study larger examples of SRG's,
and also consider other types of graphs such as the
Cai, F\"urer and Immerman
graph constructions~\cite{cai:89}, which 
are too large to be studied with the methods used here and would require
Quantum Monte Carlo simulations. 


In the current study we have distinguished between graphs by measurements
of average quantities like
$M_x=\langle \sum_i \sigma^x_i\rangle$ (i.e., magnetization along the
$x$-direction) and $\sum_{i\ne j} \langle \sigma^z_i \sigma^z_j
\rangle^2$. More information would be obtained by more sophisticated analyses of measurement data. One example, would be to calculate higher moments of the obtained data rather than using only averages as we have done in this study. Another example would be comparing the individual
(sorted) values rather than the average over sites~\cite{farhiPrivate}.
This should be done in future studies. 


It is important to understand 
the efficiency of the algorithm we propose, i.e.\ how
the running time of the algorithm scales with graph size for large sizes.  This requires
an analysis of the size dependence of the minimum gap for large sizes,
presumably using
Quantum Monte Carlo simulations,
e.g.~\cite{hen:11}, to get to large enough sizes to see the trend.


An attractive feature of the method proposed here is that it can be
implemented in principle by existing (albeit prototypical) quantum annealers.
We require a quantum annealer that has
Ising-spin interactions of finite connectivity plus constant
transverse magnetic fields. The prototypical ``D-Wave One"
machine~\cite{johnson:11} has these capabilities. 
However, D-Wave One can not
calculate the quantities studied here, $\langle \sigma^x_i\rangle$ and
$\langle \sigma^z_i \sigma^z_j\rangle $, though it can calculate averages of a
single $\sigma^z$ and also susceptibilities involving
$\sigma^z$~\cite{aminPrivate}. For the model studied here, $\langle
\sigma^z_i\rangle$ is zero because of the bit-flip symmetry of the
Hamiltonian. However, one could break this symmetry by
adding a field coupling to $\sigma^z$ and use the values of
the $\langle\sigma^z_i \rangle$ to
distinguish between graphs. In other words, it seems likely to us that
quantities which D-Wave One \textit{can} calculate would also be useful for the
Graph Isomorphism problem.  Another useful quantity which can 
be determined by D-Wave hardware~\cite{aminPrivate} is the
distribution of the energy as a function of the evolution time of the
algorithm.  This distribution would be sensitive to the details of spectrum of
the graph and is therefore also likely be a good discriminant between graphs.

\begin{acknowledgments}

We thank Kenneth Rudinger, Mohammad Amin, and Eddie Farhi for helpful
comments.  We acknowledge support from the National Security Agency (NSA)
under Army Research Office (ARO) contract number W911NF-09-1-0391, and from
the National Science Foundation under Grant No.~DMR-0906366. 
\end{acknowledgments}
\bibliography{refs,notes}

\begin{thebibliography}{47}
\bibitem[{bit()}]{bit-flip}
\bibinfo{note}{When quoting degeneracies we will neglect the fact that each
  state of $\hat{H}_p(G)$ has the same energy as the state with all bits
  flipped, because of bit-flip symmetry. The true degeneracy is therefore twice
  the value we quote.}
\end{thebibliography}

\appendix
\section{\label{app:cgm}  Conjugate gradient method}

In this appendix we explain the version of the conjugate-gradient method that was used in
this study to calculate the ground state of a given Hamiltonian.  We found
that the `traditional' Lanczos and conjugate-gradient methods are not
sufficiently accurate for our problem, so we had to develop a new
variant of these methods.

The ground state of a Hamiltonian $\hat{H}$ is obtained by minimizing
\beq
E_0=\min f(|\psi\rangle  )=\min_{\{| \psi \rangle \}}
\frac{\langle \psi | \hat{H} | \psi \rangle}{\langle \psi | \psi \rangle} \, .
\eeq
The key feature is that the objective function $f$ has no local minima
but only the one global minimum that we need. Hence, in principle, any
minimization method would work, but we find that this is not the case in
practice because of the limitations of finite-precision arithmetic.
After choosing a
basis $\{ | n \rangle \}$, one can write $| \psi \rangle=\sum c_n
|n\rangle$, and the objective function to be minimized becomes
\beq
E_0&=&\min f(\{c_n\} ) \\\nonumber 
&=& \min_{\{c_n\} }  \frac{\sum_n c_n^2 d_n +\sum_n c_n
\sum_{m} a_{nm} c_m}{\sum_n c_n^2} \,,
\eeq
in which
we denote the diagonal elements of the Hamiltonian by $\hat{H}_{nn}=d_n$ and
the off-diagonal elements (presumably sparse) by $\hat{H}_{nm}=a_{nm}$.  
In what follows we shall assume that the coefficients $c_n$ are real-valued 
for convenience. The generalization to complex-valued coefficients is trivial. 
 
The basic conjugate gradient method in this case is very simple. 
Firstly, the gradient with respect to the various parameters $c_n$ is easily obtainable:
\beq
\frac{ \partial f}{\partial c_k} &=& 
 \frac{2 d_k c_k + 2 \sum_{m} a_{km} c_m}{\sum_n c_n^2}
\\\nonumber &-& \frac{2 c_k \left( \sum_n c_n^2 d_n +\sum_n c_n \sum_{m} a_{nm} c_m\right) }{\left( \sum_n c_n^2\right)^2} \, .
\eeq
Secondly, the one-dimensional minimization steps of the conjugate gradient method can be calculated explicitly 
\beq
&\min_{\alpha}& f(|\psi\rangle + \alpha | \delta\rangle) =\\\nonumber
&\min_{\alpha}& \frac{ \langle \delta | \hat{H} |\delta \rangle \alpha^2
+2 \Re \langle \psi | \hat{H} | \delta \rangle \alpha +\langle \psi | \hat{H} | \psi \rangle }
{ \langle \delta |\delta \rangle \alpha^2
+2 \Re \langle \psi | \delta \rangle \alpha +\langle \psi |\psi \rangle} \,.
\eeq
The above expression is minimized for $\alpha = \alpha*$ which is
one of the two solutions of a quadratic equation, and this produces the
minimum (along the line) energy of:
\beq
E_0 &=&  f(|\psi\rangle + \alpha^* | \delta\rangle) \\\nonumber &=& 
\frac{ \langle \delta | \hat{H} |\delta \rangle {\alpha^*}^2
+2 \Re \langle \psi | \hat{H} | \delta \rangle \alpha^* +\langle \psi | \hat{H} | \psi \rangle }
{ \langle \delta |\delta \rangle {\alpha^*}^2
+2 \Re \langle \psi | \delta \rangle \alpha^* +\langle \psi |\psi \rangle} \,
.
\eeq

\subsection{Variable-offset minimization}
Because of the limitation of floating point arithmetic, inaccuracies may occur 
when adding terms with very different orders of magnitude.
This turns out to not affect the value of ground state energy very much, but
it does have a large effect on the 
ground-state coefficients, and hence also on ground-state
expectation values. 
To prevent this, we find it is necessary to 
ensure that the diagonal and the off-diagonal contributions to the energy
have the same order of magnitude. 

A simple way to do this is
to offset the diagonal term in each step such that the total
energy is kept equal to zero, meaning that the diagonal and off-diagonal terms
are equal in magnitude (and opposite in sign).  This is simply done by
shifting the diagonal elements $d_n \to d_n + \epsilon$, choosing the constant
$\epsilon$ in each step in such a way that the total energy (diagonal plus
non-diagonal) is zero.  From the equality: 
\beq
&&\frac{\sum_n c_n^2 (d_n+\epsilon)  +\sum_n c_n \sum_{m} a_{nm} c_m}{\sum_n |c_n|^2}
\\\nonumber
&=&\frac{\sum_n c_n^2 d_n  +\sum_n c_n \sum_{m} a_{nm} c_m}{\sum_n |c_n|^2}+ \epsilon 
\,,
\eeq
it follows that offsetting the objective function by $\epsilon$ will not
change the amplitudes $c_n$, and so will also not affect the gradients.  Its
only effect is to offset the minimum energy to zero at each step. 
The resulting ground-state energy at the end of
the process is stored in the variable offset, $E_0=\epsilon$. 

\subsection{Minimizing the residuals}
Once the energy $E_0$ is measured to great precision, we followed this up with
a second stage of minimizing the sum of squares of the residuals, where the
residuals are the elements of
$\hat{H} | \psi \rangle  - E_0 | \psi \rangle$. We therefore
perform a conjugate
gradient routine on the objective function:
\beq
f  = \min_{\{| \psi \rangle \}} \frac{ \langle \psi | \hat{H}^2 | \psi \rangle 
- 2 E_0 \langle \psi |  \hat{H} | \psi \rangle +
E_0^2 \langle \psi  | \psi \rangle}{\langle \psi  | \psi \rangle}\,,
\eeq
at the end of which $| \psi \rangle$ is a good approximation to 
the ground state wavefunction.

\section{\label{app:nonsrg} Non-SRG graphs}

Here we provide the adjacency matrices for the models
studied in Sec.~\ref{sec:nonsrg}.  
Both $G_1$ and $G_2$ are regular graphs on $14$ vertices with valency $4$ and
are found in \cite{emms:06,emms:09}.  The graph $G_1$ is given by the
adjacency matrix:
\begin{equation}
\mathbf A_{G_1} = \left(
\begin{array}{cccccccccccccc}
0&	1&	0&	0&	0&	0&	0&	1&	0&	0&	0&	0&	1&	1\\
1&	0&	0&	0&	0&	0&	0&	0&	1&	1&	0&	0&	1&	0\\
0&	0&	0&	1&	0&	0&	0&	0&	1&	1&	0&	0&	0&	1\\
0&	0&	1&	0&	0&	1&	0&	1&	1&	0&	0&	0&	0&	0\\
0&	0&	0&	0&	0&	1&	0&	0&	0&	1&	1&	0&	0&	1\\
0&	0&	0&	1&	1&	0&	1&	0&	0&	0&	1&	0&	0&	0\\
0&	0&	0&	0&	0&	1&	0&	1&	0&	0&	0&	1&	0&	1\\
1&	0&	0&	1&	0&	0&	1&	0&	0&	0&	0&	1&	0&	0\\
0&	1&	1&	1&	0&	0&	0&	0&	0&	0&	0&	1&	0&	0\\
0&	1&	1&	0&	1&	0&	0&	0&	0&	0&	1&	0&	0&	0\\
0&	0&	0&	0&	1&	1&	0&	0&	0&	1&	0&	0&	1&	0\\
0&	0&	0&	0&	0&	0&	1&	1&	1&	0&	0&	0&	1&	0\\
1&	1&	0&	0&	0&	0&	0&	0&	0&	0&	1&	1&	0&	0\\
1&	0&	1&	0&	1&	0&	1&	0&	0&	0&	0&	0&	0&	0
\end{array}
\right) \,,
\end{equation}
and the graph $G_2$ is obtained from $G_1$ by replacing the entries $(1,2)$ and
$(3,4)$ with entries $(1,3)$ and $(2,4)$ (and corresponding transposed
entries). It can be verified that $G_1$ is not isomorphic to $G_2$.

Graphs $G_3$ and $G_4$, also studied in Sec.~\ref{sec:nonsrg},
are regular graphs on $16$ vertices with valency $3$ and
can be found in Ref.~\cite{guo:10}. The graph $G_3$ is given by the adjacency matrix:
\begin{equation}
\mathbf A_{G_3} = \left(
\begin{array}{cccccccccccccccc}
0&	0&	1&	1&	0&	0&	0&	0&	0&	0&	0&	1&	0&	0&	0&	0\\
0&	0&	0&	0&	1&	1&	0&	0&	0&	0&	0&	0&	1&	0&	0&	0\\
1&	0&	0&	0&	1&	0&	0&	0&	0&	0&	0&	0&	0&	1&	0&	0\\
1&	0&	0&	0&	0&	1&	0&	0&	0&	0&	0&	0&	0&	0&	1&	0\\
0&	1&	1&	0&	0&	1&	0&	0&	0&	0&	0&	0&	0&	0&	0&	0\\
0&	1&	0&	1&	1&	0&	0&	0&	0&	0&	0&	0&	0&	0&	0&	0\\
0&	0&	0&	0&	0&	0&	0&	1&	1&	1&	0&	0&	0&	0&	0&	0\\
0&	0&	0&	0&	0&	0&	1&	0&	1&	0&	1&	0&	0&	0&	0&	0\\
0&	0&	0&	0&	0&	0&	1&	1&	0&	0&	0&	1&	0&	0&	0&	0\\
0&	0&	0&	0&	0&	0&	1&	0&	0&	0&	0&	0&	1&	1&	0&	0\\
0&	0&	0&	0&	0&	0&	0&	1&	0&	0&	0&	0&	1&	0&	1&	0\\
1&	0&	0&	0&	0&	0&	0&	0&	1&	0&	0&	0&	0&	0&	0&	1\\
0&	1&	0&	0&	0&	0&	0&	0&	0&	1&	1&	0&	0&	0&	0&	0\\
0&	0&	1&	0&	0&	0&	0&	0&	0&	1&	0&	0&	0&	0&	0&	1\\
0&	0&	0&	1&	0&	0&	0&	0&	0&	0&	1&	0&	0&	0&	0&	1\\
0&	0&	0&	0&	0&	0&	0&	0&	0&	0&	0&	1&	0&	1&	1&	0\\
\end{array}
\right) \,,
\end{equation}
and the adjacency matrix of $G_4$ is obtained from that of $G_3$ by inverting
the entries  (i.e. interchanging ones with zeros) belonging to the sub-matrix spanned by rows $(1,2)$ and columns $(3,4,5,6)$ along with the corresponding transposed entries.

\end{document}